\documentclass[english,twocolumn]{revtex4-2}
\usepackage{amsmath}
\usepackage[T1]{fontenc}
\usepackage{float}
\usepackage{amssymb}
\usepackage{graphicx}
\usepackage{esint}
\usepackage[utf8]{inputenc}



\usepackage{babel}
\begin{document}
\title{Exploring the dynamics of the Kelvin-Helmoltz instability in paraxial fluids of light}
\date{August 2023}
\author{Tiago D. Ferreira}
\thanks{Authors contributed equally}
\author{Jakub Garwo\l a}
\email{jakub.garwola@gmail.com}
\thanks{Authors contributed equally}
\author{Nuno A. Silva}
\thanks{nuno.a.silva@inesctec.pt}
\address{Departamento de Física e Astronomia, Faculdade de Ciências, Universidade do Porto, Rua do Campo Alegre s/n, 4169-007 Porto, Portugal}
\address{INESC TEC, Centre of Applied Photonics, Rua do Campo Alegre 687, 4169-007 Porto, Portugal}
\begin{abstract}

Paraxial fluids of light have recently emerged as promising analogue physical simulators of quantum fluids using laser propagation inside nonlinear optical media. In particular, recent works have explored the versatility of such systems for the observation of two-dimensional quantum-like turbulence regimes, dominated by quantized vortex formation and interaction that results in distinctive kinetic energy power laws and inverse energy cascades. 
In this manuscript, we explore a regime analogue to Kelvin-Helmoltz instability to look into further detail the qualitative dynamics involved in the transition from smooth laminar flow to turbulence at the interface of two fluids with distinct velocities. Both numerical and experimental results reveal the formation of a vortex sheet as expected, with a quantized number of vortices determined by initial conditions. Using an effective length transformation scale we get a deeper insight into the vortex formation phase, observing the appearance of characteristic power-laws in the incompressible kinetic energy spectrum that are related to the single vortex structures. The results enclosed demonstrate the versatility of paraxial fluids of light and may set the stage for the future observation of distinct classes of phenomena recently predicted to occur in these systems, such as radiant instability and superradiance.
\end{abstract}
\maketitle

\section{Introduction}

In recent years, the propagation of light in nonlinear media under the paraxial regime received increasing attention as an interesting experimental platform for emulating quantum fluid dynamics \cite{carusotto2014superfluid,glorieux2023hot}. In its essence, the idea is to leverage the formal equivalence of the mathematical models describing the propagation of a laser beam in a nonlinear medium and the non-linear Schrodinger equation (NLSE) describing the temporal evolution of wavefunction of Bose-Einstein condensates (BEC) \cite{carusotto2014superfluid}. This research topic, usually called in the literature quantum fluids of light in propagating geometries or paraxial fluids of light, exploits major advantages of light-based experimental systems and constitutes a promising line of research for quantum simulations, namely for the experimental observation of quantum-like regimes \cite{abuzarli2021blast,abobaker2022inverse,fontaine2020interferences,Ferreira_2022}.

Compared with its analogue, a paraxial light fluid emulates the dynamics of the BEC along the propagation axis of the electromagnetic field, meaning that the analogue of the fluid density evolution in time of an initial state corresponds to the electric field intensity at consecutive transversal planes. Experimentally, paraxial light fluids can be achieved in distinct optical media including resonant atomic systems \cite{glorieux2023hot,abobaker2022inverse,fontaine2020interferences}, photorefractive crystals \cite{michel2018superfluid,Ferreira_2022}, or thermo-optic liquid solutions \cite{vocke2016role,prain2019superradiant}. Although these feature distinct degrees of complexity and specifications depending on the utilized optical media, the lower cost and simpler experimental control compared with actual BECs make them an interesting solution to explore particular aspects of quantum-like phenomenology in the laboratory. Furthermore, to this higher accessibility also adds a plethora of advantages, from which we highlight the versatile state initialization using wavefront shaping techniques, the ability to assess the phase of the initial and final state using holographic reconstruction, and the enhanced control of the potential landscape using the interaction with other optical beams under cross-Kerr effects.

One of the topics that can strongly benefit from the use of these platforms is the subject of quantum turbulence. Turbulence consists of complex fluid motion dominated by chaotic behavior and is a class of phenomena ubiquitous across multiple domains of physical sciences \cite{mccomb1990physics,tsytovich1977theory,reeves2012classical}. Although an analytical description is often unfeasible, it usually features qualitative and quantitative signatures with a certain degree of universality that can be studied. Focusing on its quantum counterpart, quantum turbulence signatures depend on the dimensionality of the physical system \cite{barenghi2023types}. In specific, for two-dimensional systems, strong turbulence regimes lead to the formation of vortex structures, featuring a characteristic $k^{-3}$ power law in the incompressible kinetic energy spectrum towards smaller scales \cite{Bradley_2012}. Furthermore, the interaction between multiple vortices may also lead to an energy transfer between smaller to large scales - an inverse energy cascade - featuring a characteristic $k^{-5/3}$ power law towards larger scales \cite{Bradley_2012}. Regarding its experimental observation, paraxial fluids of light can leverage the easy state initialization and access to phase distributions to probe theoretical predictions and provide better insight into the underlying physical phenomena. Recent work on this subject has focused mostly on the generation and observation of vortex structures past obstacles \cite{Ferreira_2022} or forced by two-wave interference and dark soliton decay \cite{abobaker2022inverse}.

Another topic of interest for quantum turbulence is the observation and exploration of the hydrodynamic instabilities which govern the transition from smooth laminar flow to chaotic turbulence. In this domain, the Kelvin-Helmholtz instability (KHI) is perhaps the most recognizable class of hydrodynamic instability and simply describes the instability formed at the interface of two streams moving with distinct relative velocities. In particular, under convenient velocity differences (for a single component fluid it occurs at all velocity values), the streams start to roll up at the interface, and the unstable perturbations start to grow exponentially, leading to turbulent phenomena. The observation of the KHI in superfluid and BECs has been suggested multiple times in the literature. Focusing on the case of BECs, KHI was studied both in single \cite{Baggaley_2018,Giacomelli_2023} and binary BECs \cite{kobyakov2014turbulence,suzuki2010crossover}, leading to the observation of vortex sheets, i.e. collections of point vortices along a specific curve. Yet, the experimental realization of the theoretical studies can be quite challenging.
 
The difficulty to explore KHI experimentally with BECs sets the stage and opportunity for this manuscript that aims to use paraxial fluids of light for such purpose. To this end, we first introduce a theoretical model for light propagating inside a photorefractive crystal, highlighting the analogy with fluids and BECs. We then introduce an experimental setting for the observation of KHI, analyzing the numerical and experimental results obtained for distinct values of analogue fluid velocity. Finally, we verify the experimental observation of qualitative and quantitative behavior in terms of vorticity and incompressible kinetic energy spectrum, which align with the expected theoretical predictions.


\section{Physical model}

\begin{figure*}
\begin{center}
\includegraphics[width=1.0\textwidth]{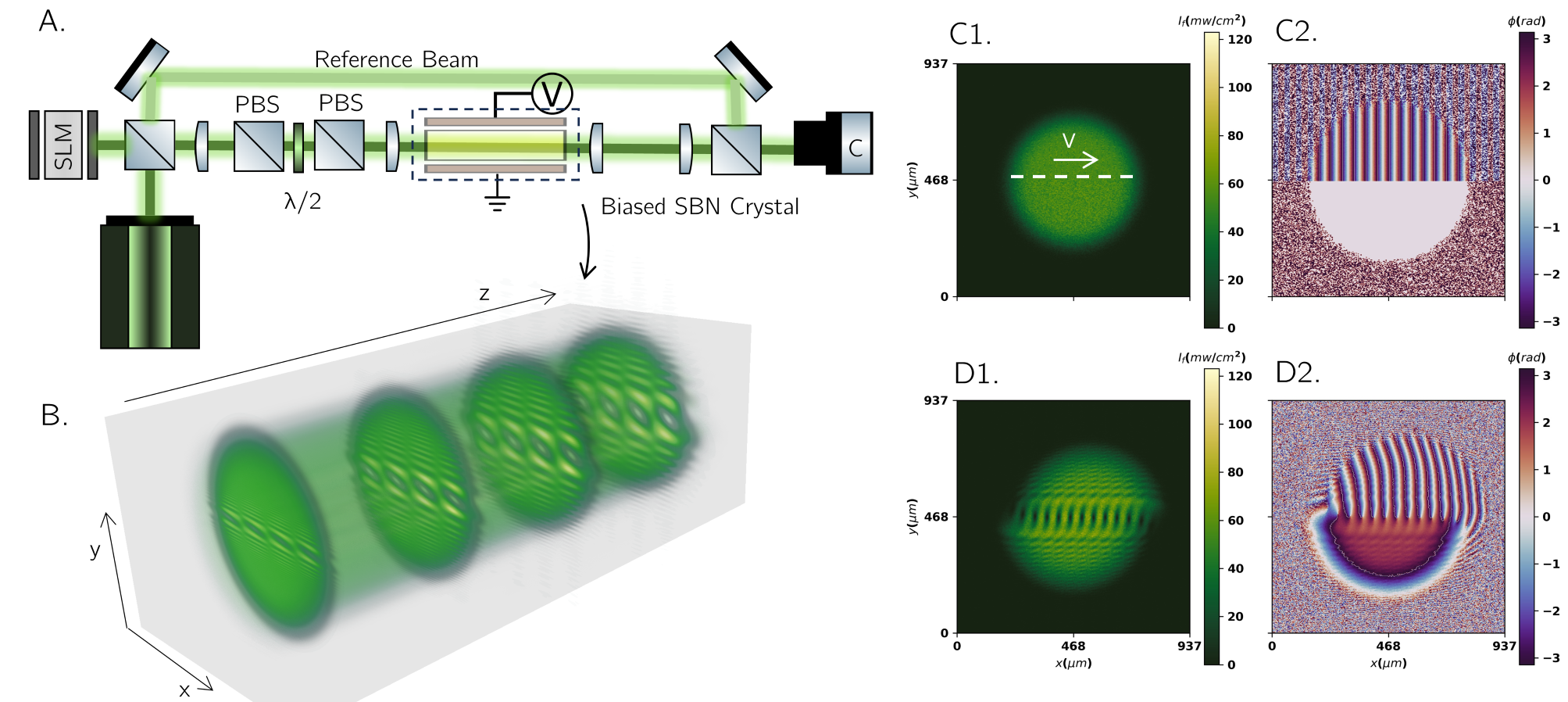}
\end{center}
\caption{Illustration of the typical experimental scenario for the observation of Kelvin-Helmoltz instability in paraxial fluids of light. A. Simplified schematic of the experimental setup. B. Illustration of the optical intensity for a typical, obtained using numerical simulations for an initial state with $v\approx4.4 c_s$ under the conditions described in the main text. C. Optical state at the input (numerical state, used for simulations, with 1. Intensity and 2. Phase) and its characteristic phase imprinted on the top half-plane using wavefront shaping techniques; D. Typical state at the output, revealing the formation of a vortex sheet along the discontinuity interface.}
\label{experimental_setup}
\end{figure*}

The main objective of this work is to explore the dynamics of Kelvin-Helmoltz instability (KHI) in paraxial fluids of light for the observation of quantum-like turbulence signatures. For this, we first focus on the dynamics of a laser beam propagating inside a photorefractive crystal, as illustrated in Figure \ref{experimental_setup}. 

Assuming a continuous wave beam and a stationary optical response regime of the crystal, the laser propagating along the axis $z$ can be expressed mathematically by $\boldsymbol{E}_f(\boldsymbol{r}_\perp, z)=E_f(\boldsymbol{r}_\perp,z)\exp{[i (\boldsymbol{k}_f z)]}\boldsymbol{e}_p$, with $E_f(\boldsymbol{r}_\perp,z)$ being the envelope function and $k_f=2\pi/\lambda_f$ being the vacuum wavenumber with $\lambda_{f}=532nm$. Considering the polarization $\boldsymbol{e}_p$ aligned with the $c-$ axis of the photorefractive crystal, and neglecting the anisotropic response of the crystal and its absorption, we can, assuming the paraxial approximation, get that the propagation dynamics of the envelope functions of the optical beam are described by \cite{boughdad_omar_2020_4404239}
\begin{equation}
i \partial_z E_f + \frac{1}{2n_ek_f} \nabla_\perp^2 E_f - k_f\Delta n_{max} \frac{|E_f|^2}{|E_f|^2+ I_{sat}} E_f = 0,
\label{eq:fluid_eq}
\end{equation}
where $n_e$ is the extraordinary refractive index, $I_{sat}$ is the saturation intensity, and $\Delta n_{max}=1/2n_e^3r_{33}E_{ext}$\cite{boughdad_omar_2020_4404239}. For the sake of simplicity, we can introduce coordinate transformations $x'=k_f\sqrt{\Delta n_{max}n_e}x$ and $z'=k_f\Delta n_{max}z$, which dropping the primes, leads to 
\begin{equation}
i \partial_z E_f + \frac{1}{2} \nabla_\perp^2 E_f - \frac{|E_f|^2}{|E_f|^2+ I_{sat}} E_f= 0.
\label{eq:fluid_equation_n1}
\end{equation}
For regimes far away from saturation, i.e. $I_f = |E_f|^2 \ll I_{sat}$, the model can be approximated by
\begin{equation}
i \partial_z E_f + \frac{1}{2} \nabla_\perp^2 E_f - \frac{1}{I_{sat}} |E_f|^2 E_f= 0.
\label{eq:fluid_equation_n}
\end{equation}

\subsection{Quantum Fluid Analogy}

Looking at equation \ref{eq:fluid_equation_n} the analogy between the dynamics of light inside the photorefractive crystal and BECs becomes clear. Indeed, in mean-field theory assuming binary collisions, the Schrödinger equation for the condensate wavefunction $\psi(\vec{x},t)$ takes the form of a Nonlinear Schrödinger equation
\begin{equation}\label{shrodinger}
i\hbar\frac{\partial\psi(\vec{x},t)}{\partial t}=[-\frac{\hbar^{2}\nabla_{\perp}^{2}}{2m}+V(\vec{x},t)+g|\psi(\vec{x},t)|^{2}]\psi(\vec{x},t),
\end{equation}
where $m$ is the mass of the particle, $g$ is a coupling constant related to the binary collisions and $V(\vec{x},t)$ an additional external potential. Due to the mathematical equivalence of both models, it is easy to support that the analogue of temporal dynamics of a BEC can be observed by looking at multiple planes taken along the propagation direction of the crystal \cite{silva2021, boughdad_omar_2020_4404239,glorieux2023hot,Ferreira_2022,carusotto2014superfluid}. It is interesting to notice other formal relations: the quantum fluid density $|\psi_f|^2$ can be related to the optical intensity $I_f=|E_f|^2$, the interaction via particle collision becomes an interaction mediated by the nonlinear optical properties, and the atomic mass has its analogue as an effective mass through diffraction given by $k_f$.

Furthermore, to reinforce the concept of paraxial fluid of light and in particular unravel what is the equivalent of the fluid velocity, we can consider the equation \ref{eq:fluid_equation_n1} and apply the Madelung transformation 
\begin{eqnarray}
\nonumber
\psi_f = \sqrt{I_f(\boldsymbol{r}_{\perp}, z)} e^{i \phi(\boldsymbol{r}_{\perp}, z)} 
\end{eqnarray}
where $\phi$ is the spatial phase distribution of the optical beam. By substitution, and separating the imaginary and real part, we can obtain a set of Navier-Stokes equations
\begin{eqnarray}
\frac{\partial I_f}{\partial z} + \nabla_\perp \cdot \left( I_f \boldsymbol{v} \right) = 0,
\label{eq:hydro_1}
\end{eqnarray}
\begin{eqnarray}
\frac{\partial \boldsymbol{v}}{\partial z} + \left( \boldsymbol{v} \cdot \nabla_\perp\right) \boldsymbol{v} = \nabla_\perp\left(\frac{I_f}{I_{sat}} + \frac{\nabla_\perp^2 \sqrt{I_f}}{2\sqrt{I_f}}\right).
\label{eq:hydro_2}
\end{eqnarray}
by considering that the fluid velocity is related to the phase spatial distribution as $\boldsymbol{v} = \nabla_\perp \phi$, meaning that one can control it experimentally by imprinting a given phase profile using, for example, a spatial light modulator. The last term on the right side of equation \ref{eq:hydro_2} is known as Bohm potential, known to be related to quantum-like effects. 

Straightforward linearization of equations \ref{eq:hydro_1} and \ref{eq:hydro_2} using\cite{Chiao_1999}
\begin{eqnarray}
I=I_0+\delta I
\end{eqnarray}
\begin{eqnarray}
v=v_0+\delta v
\end{eqnarray}
leads to the Bogoliubov dispersion relation for elementary excitations on top of the photon fluid with intensity $I_0$, which in the laboratory reference frame becomes
\begin{eqnarray}
k_z=\sqrt{\frac{k_{\perp}^2}{2k_fn_e}\left(\frac{k_{\perp}^2}{2k_fn_e}+2k_f\Delta n\right)}. 
\label{eq:bogolioubov}
\end{eqnarray}
From this, follows the analog sound velocity of the fluid
\begin{eqnarray}
c_s=\sqrt{\frac{\Delta n_{max} I_f}{n_e I_{sat}}}, 
\label{eq:sound_v}
\end{eqnarray}
and the healing length of typical length scales in the transversal plane
\begin{eqnarray}
\xi=\frac{1}{k_f\sqrt{n_e\Delta n_{max} I_f/I_{sat}}}.
\label{eq:healing}
\end{eqnarray}

\subsection{Kelvin-Helmoltz Instability and the onset of quantum turbulence}

In general, Kelvin-Helmoltz instability (KHI) describes the instability formed at the interface of two streams moving with distinct relative velocities. Occurring at all velocity values for a single fluid, in its classical version the two streams start to roll up at the interface, leading to unstable perturbations that induce a turbulent regime.

In the quantum version of this effect, proposed for BECs in multiple references in the literature (e.g. \cite{Giacomelli_2023,baggeley2018,kokubo}), a line of quantized vortices is seeded at the interface between the fluids. For this case, and due to the irrotationality of the velocity vector field, the condensate enters in a turbulent regime featuring an array of vortices along the interface, mimicking the roll-up observed on the classical KHI.

To be more specific, we focus on the case of an initial flat-top state,
\begin{equation}\label{initial_state}
    E(r_{\perp},z=0) = \sqrt{I_0} \exp {\left[-2\frac{r^2_{\perp}}{w^2}\right]^{4}} \exp {\left({iv_{x}x\theta(y)}\right)},
\end{equation}
with a waist size $w$, and the gradient in the phase defining an analogue fluid velocity $v_x$. For this case, the vortices are seeded in the points where the phase difference between the bottom and top half is $\pi$, due to instabilities related with phase discontinuity\cite{baggeley2018}. For this reason, they appear at regular spatial intervals \cite{Giacomelli_2023}
\begin{eqnarray}
\delta x=\frac{2\pi}{v_x},
\label{eq:vortex_period}
\end{eqnarray}
and therefore, for a supergaussian of waist $w$ one shall expect the formation of $n_v = floor(w/\delta x)$ vortices. 

Although for longer propagation distances (equivalent to simulation time) the vortices can progressively aggregate into larger clusters \cite{Sitnik_2022}, the single vortex approximation \cite{Bradley_2012} is sufficient to understand some of the signatures of this regime for shorter propagation distances. Assuming the vortices to be independent solutions of the Gross-Pitaevskii Equation, they take the form
\begin{equation}\label{single_vortex_function}
 E_v(r_{\perp},z)=\sqrt{I_{0}}\chi(r_\perp/\xi)e^{-i  n_e k_f z}e^{il\theta}.
\end{equation}
with $l\in\mathbb{Z}$. The vortex core structure is described by the radial function $\chi(r_\perp/\xi)$, which indicates that the size of the vortex is fixed by the healing length \cite{Bradley_2012}. Then it is straightforward to obtain the velocity as 
\begin{equation}
\vec{v_{\theta}}=\frac{l}{r}\vec{e}_{\theta},
\end{equation}
and subsequently
\begin{equation}
\vec{\omega}=\nabla\times\vec{v}=l\delta (r_\perp) \hat{z},
\label{vorticity}
\end{equation}
meaning that the vorticity is localized in the middle of the vortex and its value is quantized. An interesting signature of the presence of vortices, and related to strong turbulence dynamics, is the characteristic incompressible kinetic energy spectra of the vortex field. This quantity can be obtained from the density-weighted velocity
\begin{equation}
    \vec{u}(r_{\perp},t) = |\psi(r_{\perp},z)| \cdot \vec{v}(r_{\perp},z) = \vec{u^i}(r_{\perp},z) + \vec{u^c}(r_{\perp},z)
\end{equation}
where the components
\begin{align}
    \nabla \cdot \vec{u^i}(r_{\perp},t) &= 0 \\
    \nabla \times \vec{u^c}(r_{\perp},t) &= 0.
\end{align}
refer to the incompressible and compressible parts of the density-weighted velocity.
The incompressible kinetic energy will then correspond to the rotational energy of vortices as
\begin{equation}
    E^i = \frac{1}{2} \int d\vec{r} \left| \vec{u^i} (r_{\perp},t) \right|^2.
\end{equation}
To obtain the dependency of $E^i$ with respect to the modulus of the momentum and thus further investigate the spectra, we perform the Fourier transform of $\vec{u^i}$ in the transverse direction and integrate on the angular coordinate. As it satisfies  $\vec{k} \cdot \vec{u^i} = 0$ we have
\begin{equation}
    u^i_\alpha (\vec{k},t) = \left( \delta_{\alpha \beta} - \frac{k_\alpha k_\beta}{k^2} \right) u_\beta(\vec{k},t) ,
\end{equation}
where lower indexes indicate vector components and the summation convention was adopted, leading to
\begin{equation}
    E^i (k) = \frac{k}{2} \int_0^{2 \pi} d \phi_k \left| \vec{u^i} (\vec{k},t) \right|^2.
\end{equation}
In the single vortex approximation (SVA), it can be shown \cite{Bradley_2012} that the low and high energy asymptotic behavior of $E^i(k)$ is given respectively by 
\begin{align*}
    E^i(k) \big|_{k\xi\ll1} &= \frac{\Omega \xi^3}{k\xi}, \\ E^i(k) \big|_{k\xi\gg1} &= \chi'(0)^2 \frac{\Omega \xi^3}{(k\xi)^3}.
\end{align*}
We note that this low energy power law is valid for systems of multiple vortexes only when they are well separated.


\begin{figure*}
\begin{center}
\includegraphics[width=1.0\textwidth]{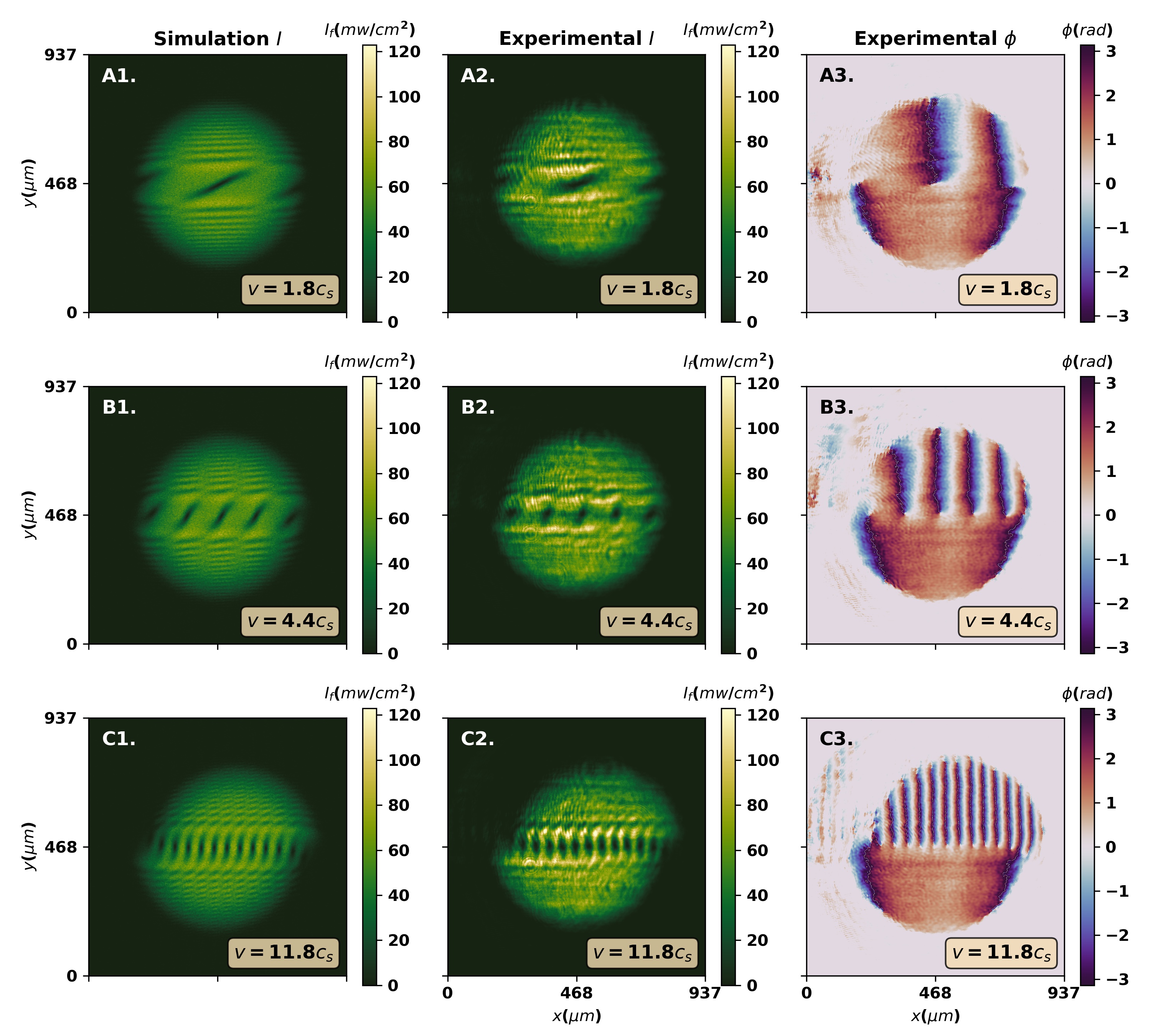}
\end{center}
\caption{Profiles obtained using numerical(Column 1) intensity field) and experimental(Column 2 and Column 3 for intensity field and phase distribution) methodologies for the output states obtained for distinct velocities (A-C) as shown in the picture, generating a different number of vortices at the interface.}
\label{profile_vx}
\end{figure*}

\begin{figure}
\begin{center}
\includegraphics[width=0.40\textwidth]{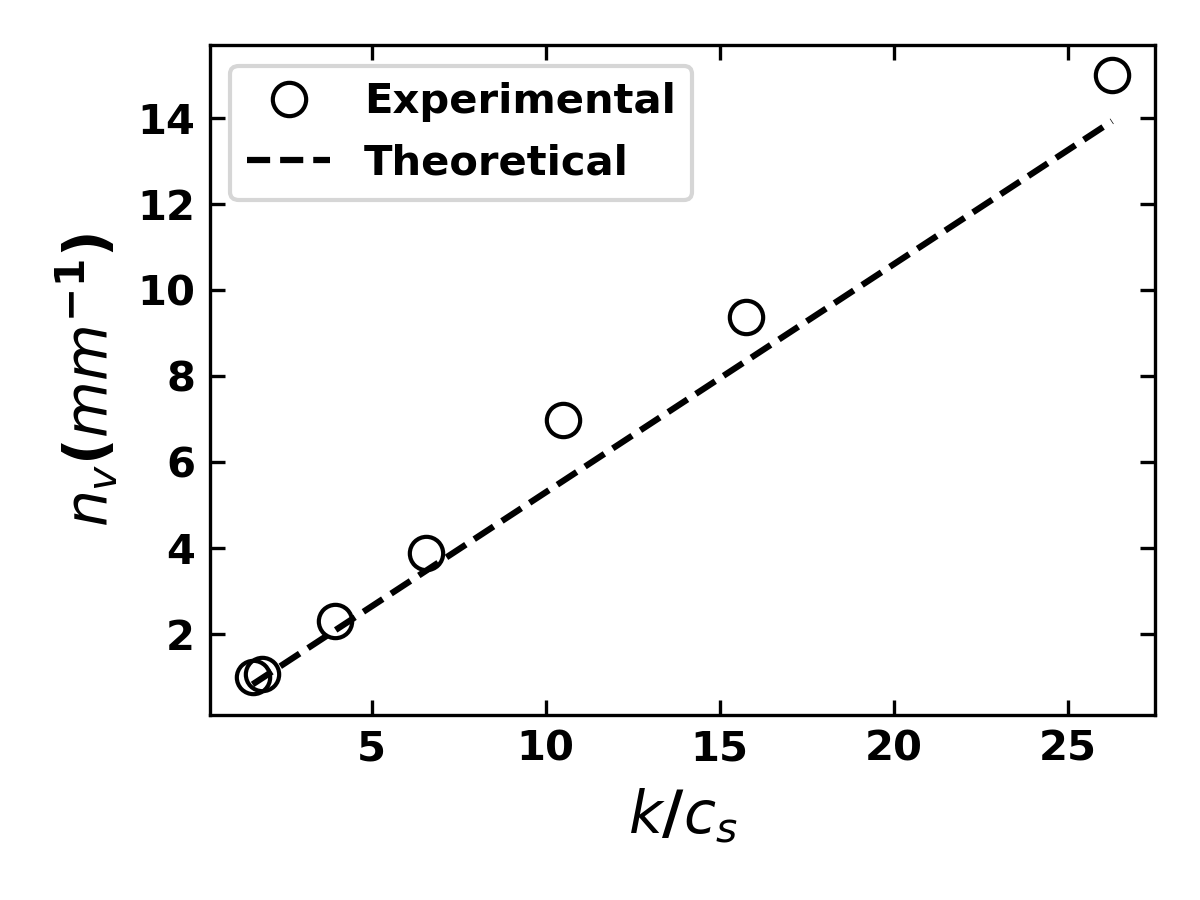}
\end{center}
\caption{Plot comparing the observed the linear vortex density $n_v = N_{vortices}/w$(circles) and the theoretical(dashed line) predicted value $n_v = (v_x w )/(2\pi)$ as described in the main text.}
\label{vortex_density}
\end{figure}

\section{Results}
Using the flat-top state introduced in equation \ref{initial_state} with waist $w=400\mu m$,  $\lambda = 532 \space nm$, and intensity $I_0\approx55 mWcm^{-2}$, we explored the regime of KHI for distinct velocities $v_x$ using both experimental methods and numerical results.

\subsection{Experimental setup}
For the experimental setup, we used a SBN:61 photorefractive crystal with dimensions $5 \times 5 \times 20 \space \text{mm}^3$ as the nonlinear medium, doped with cerium at $0.002 \%$  to increase its photorefractive effect. The crystal with refraction index $n_{e} = 2.36$ was further biased with a static electric field  $E_{0} = 8\times10^{4}V/m$, resulting on an refraction index variation $\Delta n_{max}=\frac{1}{2}n_{e}^{3}r_{33}E_{0} \sim 1.25\times10^{-4}$ assuming $r_{33} = 250\times10^{-12} pm/V$ and neglecting the absorption effects. Additionally, a white incoherent light was used to adjust the crystal $I_{sat}$ around the $450mWcm^{-2}$.

The analogue fluid velocity is controlled by imprinting the necessary phase gradients through a Spatial Light Modulator. The beam is then imaged at the input of the crystal with a 4f-system and imaged and magnified at the output with another 4f-system. The profile and phase of the output are reconstructed using a CMOS camera and an off-axis holographic technique \cite{Verrier:11} which allows retrieving the phase profile of the output beam and access to further quantities such as the angular momentum, vorticity, and kinetic energy.


Finally, one limitation of these setups is that it is impossible to assess intermediate states, i.e. distinct transversal planes inside the crystal, by simply changing the imaging plane due to the nonlinear optical properties of the medium. Nevertheless, it is still possible to recreate the evolution of the fluid inside the crystal by performing a series of experiments with different beam intensities $I_0^{'} = f I_0$ \cite{abuzarli2021blast}. Indeed, it can be shown that this transformation effectively rescales the system along the propagation and transverse directions as 
\begin{align*}
    r_\perp' &= \sqrt{f} r_\perp, \\
    z' &= f z. \\
\end{align*}
As a result, it is possible to assess the beam at the exit of the crystal for different effective propagation lengths and thus reconstruct some dynamics of the beam.

\subsection{Numerical results}

In order to validate the results obtained at the experimental level and further understand the inner workings of the KHI in paraxial light fluids, we also performed numerical simulations by numerical integration of the equation \ref{eq:fluid_equation_n} using a standard beam propagation method \cite{tiago_artigo_solver,nuno_solver_artigo} .

For each simulation, we used the initial state given by equation \ref{initial_state} varying the input velocity $v_x$ of the top half of the flat-top beam. Using standard numerical libraries and post-processing numerical routines we computed quantities such as angular momentum, vorticity, and incompressible kinetic energy spectrum for the light fluid, to compare with the experimental results.

\subsection{Results for the KHI mechanism}

Figure \ref{profile_vx} shows typical numerical and experimental results obtained for the evolution of the flat-top state given in equation \ref{initial_state} with distinct velocities $v_x$ in terms of the predicted analogue sound velocity $c_s$. It is straightforward to observe a good qualitative agreement between numerical and experimental data, with vortices being nucleated at the interface between the top (moving) and bottom (still) half of the flat-top state. Additionally, one can also observe the generation of shock waves propagating on top of the fluid, although the anisotropy of the crystal \cite{boughdad_omar_2020_4404239} and additional noise makes this signature much less clear in the experimental results compared against the numerical ones.

It can also be seen that as expected, each vortex is being nucleated at the spatial points where the phase difference is $\pi$, meaning that the system follows the expected periodicity for vortices. This can be confirmed by the numerical comparison between the linear density of vortex defined as $n_v = N_{vortices}/w$ along the interface and the expected density $n_v = w/\delta = (v_x w)/(2\pi)$. 

\begin{figure*}
\begin{center}
\includegraphics[width=1.0\textwidth]{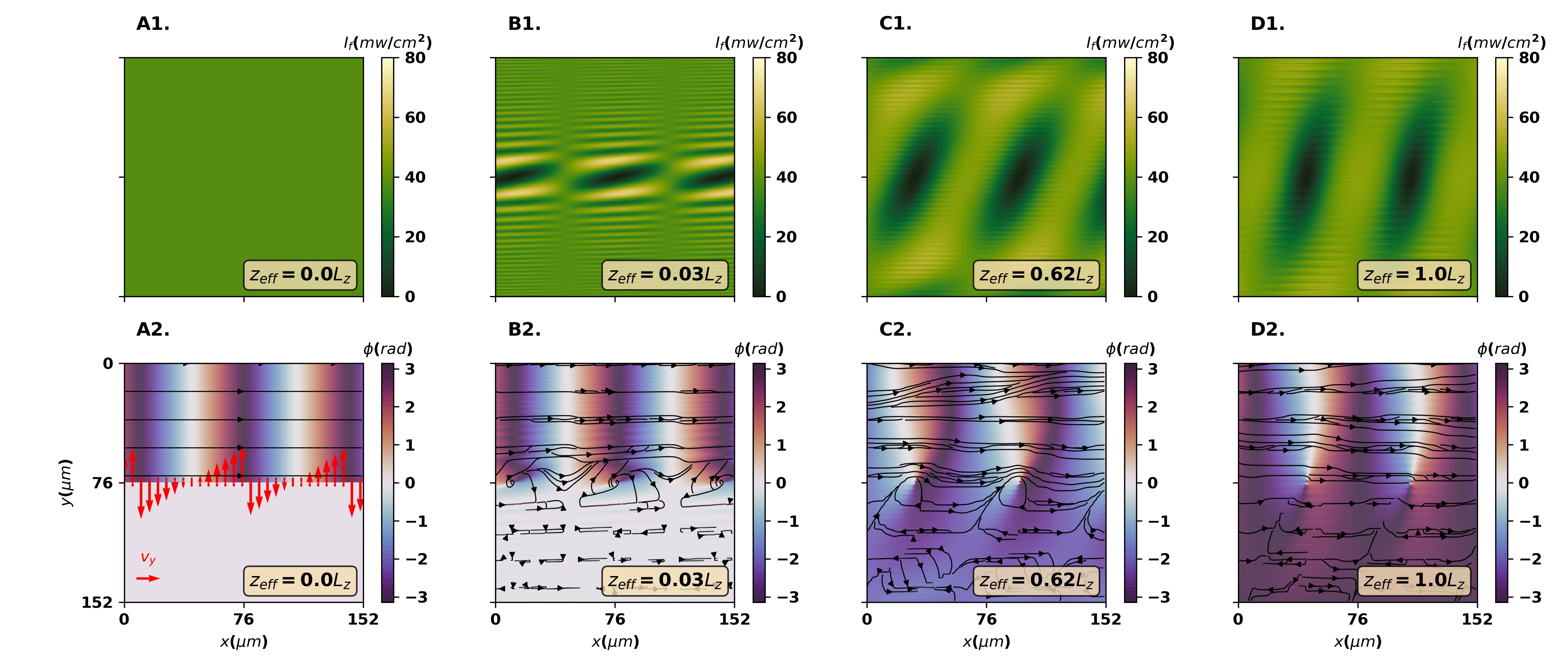}
\end{center}
\caption{Numerical results obtained for a state with velocity $v_x = 4.4 c_s$, with panels A-D representing slices taken along the propagation direction of the intensity(top) and phase(bottom) profiles. On the bottom, the streamplot depicts the gradient of the phase, associated with the analogue velocity of the paraxial fluid of light. Red Arrows in Subfigure A2 represent the $y$ component of the velocity occurring due to the discontinuity at the half-plane interface.}
\label{profile_vs_z}
\end{figure*}

To further understand the dynamics of vortex nucleation under the KHI, we first performed numerical simulations to analyze the behavior at distinct planes in the propagation distance. As it can be inferred from Figure \ref{profile_vs_z}, the phase distribution of the top half of the flat-top state creates an effective velocity component along the $y$ axis at the interface, depicted in the subfigure A2 of figure \ref{profile_vs_z} in red arrows. Focusing on the central point with phase difference $\pi$, spatial points to the left at the interface feature an upward velocity contribution while points to the right feature a downward one. Due to this mechanism, the KHI starts to develop, eventually turning into a vortex as the phase and intensity pattern starts to roll up assisted by convective currents and pressure related to the self-defocusing optical nonlinearity as described by equation \ref{eq:fluid_equation_n1}. Besides, we can also note that the same phase difference is also responsible for the presence of shock waves moving outwards, which can be better observed in Figure B1. and that also feature a degree of periodicity related with similar reasoning to the mechanism generating the KHI. 


\begin{figure*}
\begin{center}
\includegraphics[width=1.0\textwidth]{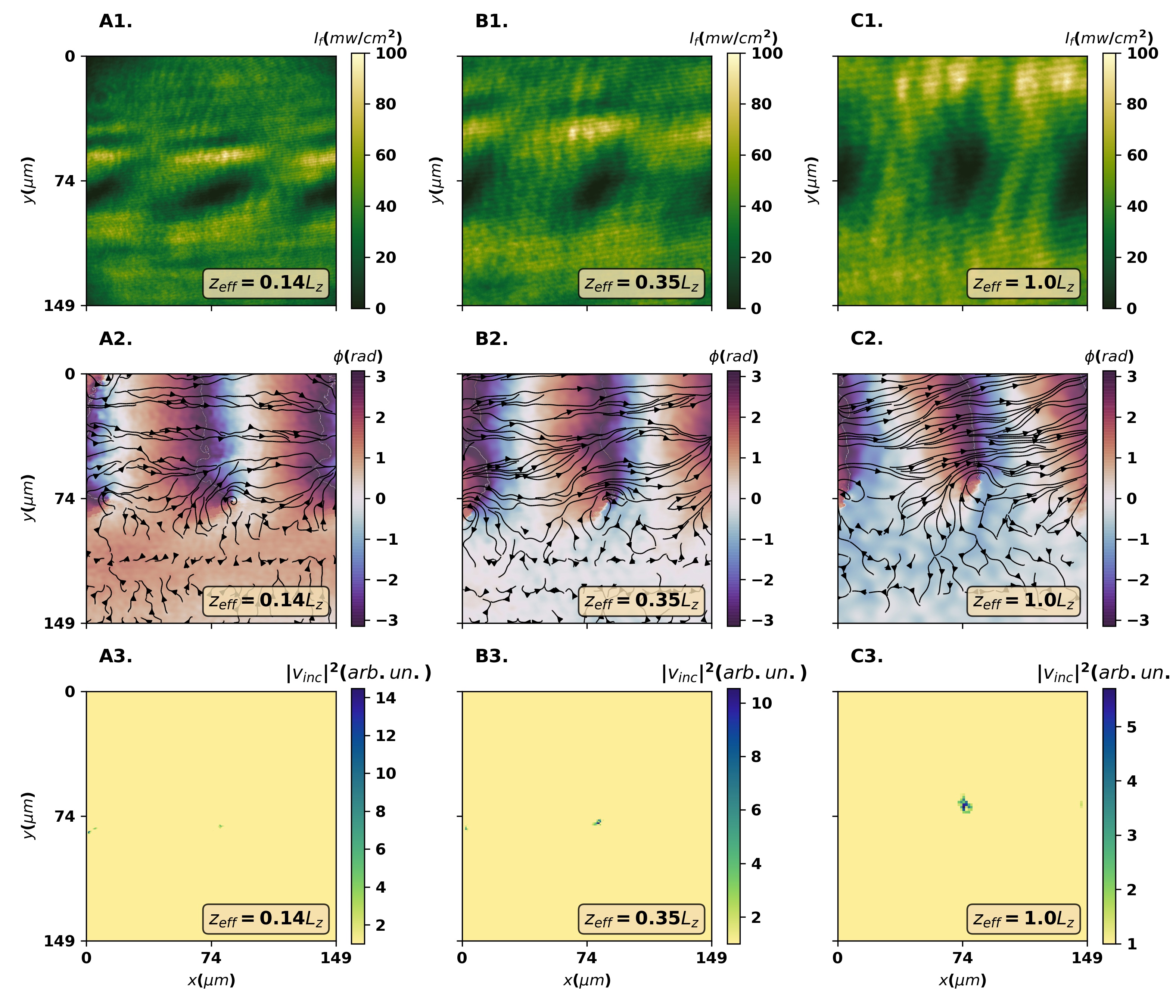}
\end{center}
\caption{Experimental results obtained for a state with velocity $v_x = 4.4 c_s$, with panels A-C representing slices taken along the propagation direction of the intensity(top) and phase(bottom) profiles using the effective scale transformation technique described in the main text. On the bottom, the set of panels A3-C3 represents the incompressible part of the velocity related to the presence of vortex structures in the velocity field, revealing the appearance of a vortex at the expected point.}
\label{effective_length}
\end{figure*}

Using the transformation of coordinates trick introduced in section III.A - which utilizes the variation of the total intensity of the input beam to change the effective length of the sample, thus allowing to reconstruct of the evolution of the fluid without accessing the inside of the crystal - we obtained the experimental results presented in figure \ref{effective_length}. Again, the results qualitatively match those obtained numerically and depicted in Figure \ref{profile_vs_z}. Furthermore, by computing the incompressible part of the velocity, one can also observe that non-zero regions are related with the spatial points where the vortices form, as observed in the set of panels A3-C3 of figure \ref{effective_length}.

\begin{figure}
\begin{center}
\includegraphics[width=0.45\textwidth]{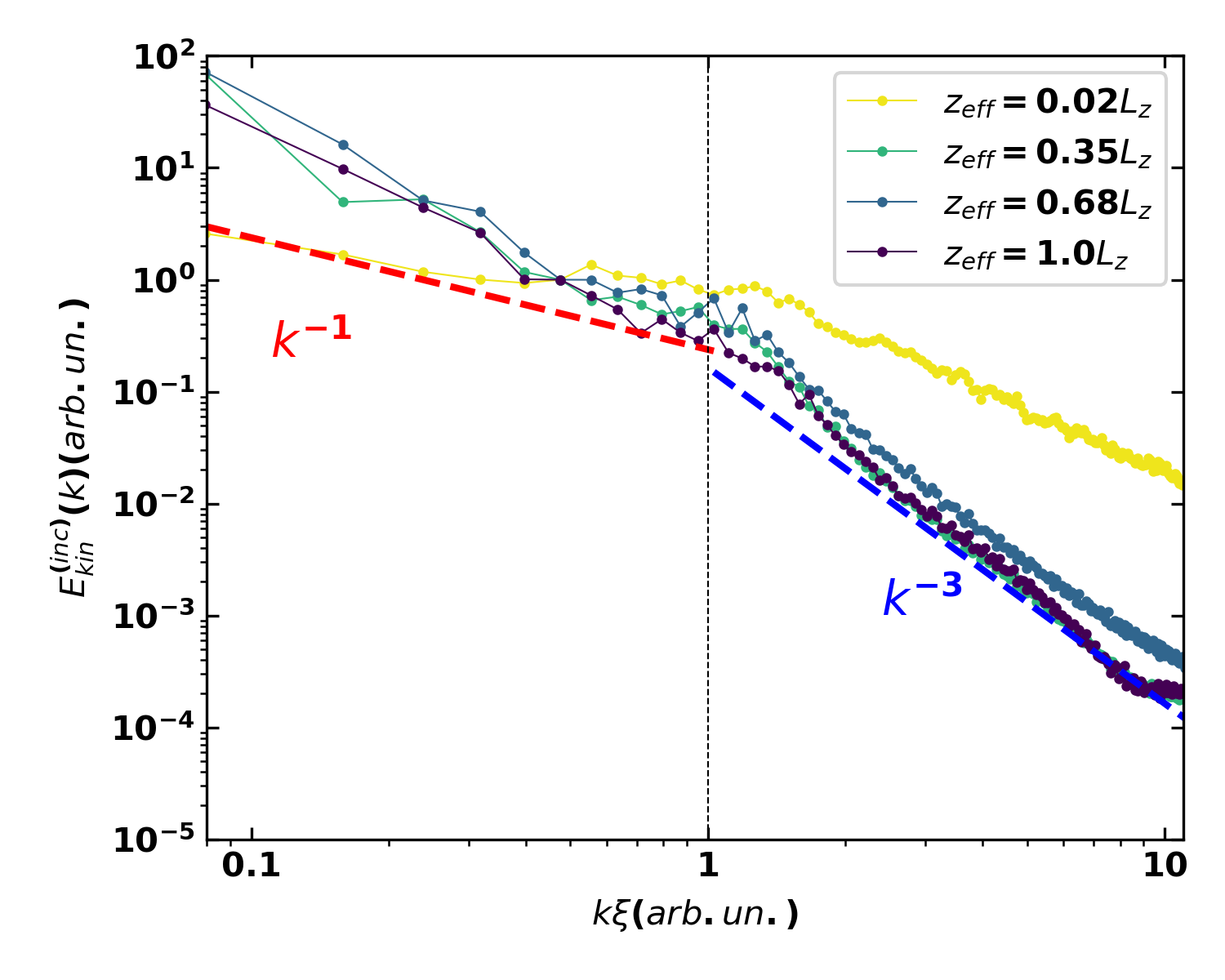}
\end{center}
\caption{Plot showing the incompressible energy spectrum in arbitrary units obtained for distinct effective propagation lengths, confirming the presence of the expected power laws for the single vortex distribution, i.e. $k^{-1}$ in clear contrast with the $k^{-3}$ power law observed towards smaller scales as the instability develops into a vortex.}
\label{kinetic_energy}
\end{figure}

Transforming the incompressible part of the kinetic energy to the Fourier space, one can also seek for the emergence of the characteristic power law for the single vortex in the incompressible kinetic energy spectrum, featuring a transition to a $k^{-1}$ (for $k \xi \ll 1$) and $k^{-3}$ (for $k \xi \gg 1$) power law after some propagation distance as predicted\cite{Bradley_2012}. Indeed, the results obtained in figure \ref{kinetic_energy} match the theoretical predictions \cite{Bradley_2012,abobaker2022inverse}. For small propagation lengths, phase slips dominate the spectra, and no characteristic power-law appears in the incompressible kinetic energy spectra as the states do not have well-defined vortices at smaller scales. Entering larger effective propagation distances, the single vortexes start to develop, which can be observed both qualitatively in the profile and phase spatial distributions, but also quantitatively, with the appearance of the characteristic power laws expected to occur. This signature confirms the onset of a vortex turbulence regime after KHI.

\section{Concluding remarks}

In this manuscript, we investigated the dynamics of an analogue of a Kelvin-Helmholtz instability in paraxial fluids of light and reported its observation, aiming to provide a better understanding of this phenomenology and establish a parallel to that expected to be observed in two-dimensional quantum fluids\cite{Giacomelli_2023}. By making use of an experimental setup with a photorefractive crystal and optical phase control using wavefront shaping techniques, we experimentally observed an analogue behavior to the KHI regime in the interface of a fluid featuring distinct transverse velocities. Qualitatively and as expected, the results obtained demonstrate the nucleation of a vortex sheet along the velocity discontinuity interface. 

The nucleation of these vortices is found to occur at specific points having phase discontinuity equal to $\pi$ between the bottom (still) and top (moving with velocity $v$) half-planes, which is validated through the comparison with the expected linear vortex density, in agreement with the theoretical predictions. The instability and vortex generation mechanism is further investigated by utilizing effective coordinate transformations using the variation of the power beam, as previously explored in the literature\cite{abuzarli2021blast}. We then establish a connection between this KHI regime and a vortex turbulence regime, confirming it quantitatively with the observation of characteristic $k^{-1}$ and $k^{-3}$ power laws in the incompressible kinetic energy spectrum. All the results presented align with theoretical predictions and those obtained using numerical methods, confirming the potential of the experimental setup presented here for the exploration of instability and turbulence signatures of two-dimensional quantum fluids of light.

Finally, the current experimental configuration allows for the exploration of various topologies, such as those with periodic boundary conditions. This could serve as the foundation for future research into, for instance, the observation of superradiance signatures in such systems \cite{Giacomelli_2023,Baggaley_2018}.

\section*{Acknowledgments}
This work is financed by National Funds through the Portuguese funding agency, FCT – Fundação para a Ciência e a Tecnologia, within project UIDB/50014/2020. T.D.F. is supported by Fundação para a Ciência e a Tecnologia through Grant No. SFRH/BD/145119/2019. N.A.S. also acknowledges the financial support of the project “Quantum Fluids of Light in Hot Atomic Vapors”, supported by FCT in collaboration with the Ministry of Education, Science and Technological Development of the Republic of Serbia.


\bibliography{bib}

\end{document}